\documentclass{webofc}
\usepackage[varg]{txfonts}   

\usepackage{comment}
\usepackage{datetime2}

\usepackage[colorlinks=true, citecolor=blue, linkcolor=blue, urlcolor=blue, pdfborder={0 0 0}, breaklinks=true]{hyperref} 
\renewcommand{\url}[1]{\href{#1}{#1}}
\renewcommand{\doi}[1]{\url{https://doi.org/#1}}

\begin{document}

\title{Integrating LHCb workflows on HPC resources: \\status and strategies} 

\author{
\firstname{Federico} 
\lastname{Stagni}\inst{1}
\fnsep\thanks{\email{federico.stagni@cern.ch}
} 
\and
\firstname{Andrea}
\lastname{Valassi}\inst{2}
\and
\firstname{Vladimir}
\lastname{Romanovskiy}\inst{3}
}

\institute{CERN, EP Department, 
Geneva, Switzerland
\and
CERN, IT Department, 
Geneva, Switzerland
\and
NRC Kurchatov Institute", IHEP, 
Protvino, Russia
}

\def\lhcb   {\mbox{LHCb}\xspace}
\def\belle  {\mbox{Belle2}\xspace}
\def\clic   {\mbox{the Linear Collider}\xspace}
\def\cta    {\mbox{Cherenkov Telescope Array}\xspace}
\def\cern   {\mbox{CERN}\xspace}
\def\lhc    {\mbox{LHC}\xspace}
\def\mc     {\mbox{Monte Carlo}\xspace}

\def\dirac      {\mbox{\textsc{DIRAC}}\xspace}
\def\lhcbdirac  {\mbox{\textsc{LHC}b\dirac}\xspace}
\def\ganga      {\mbox{\textsc{Ganga}}\xspace}
\def\python     {\mbox{\textsc{Python}}\xspace}
\def\git        {\mbox{\textsc{GIT}}\xspace}

\abstract{
High Performance Computing (HPC) supercomputers 
are expected to play an increasingly important role 
in HEP computing in the coming years. 
While HPC resources are not necessarily 
the optimal fit for HEP workflows, 
computing time at HPC centers on an opportunistic basis 
has already been available to the LHC experiments for some time, 
and it is also possible that part 
of the pledged computing resources will be offered 
as CPU time allocations at HPC centers in the future.
The integration of the experiment workflows 
to make the most efficient use of HPC resources 
is therefore essential.
This paper describes the work that has been necessary 
to integrate LHCb workflows at a specific HPC site,
the Marconi-A2 system at CINECA in Italy,
where LHCb benefited from a joint PRACE (Partnership for Advanced Computing in Europe) allocation
with the other Large Hadron Collider (LHC) experiments.
This has required addressing two types of challenges: 
on the software application workloads,
for optimising their performance
on a many-core hardware architecture
that differs significantly 
from those traditionally used in WLCG (Worldwide LHC Computing Grid), 
by reducing memory footprint 
using a multi-process approach;
and in the distributed computing area, 
for submitting these workloads using 
more than one logical processor per job,
which had never been done yet in LHCb.
}
\maketitle

\section{Introduction}
\label{sec:intro}

\lhcb is constantly looking for ways 
to opportunistically expand its distributed computing resources,
beyond those pledged by the sites of the
Worldwide LHC Computing Grid (WLCG).
One way of doing so is by integrating 
High Performance Computing (HPC) supercomputers in the \lhcb grid,
managed via \dirac~\cite{DIRAC-SW} 
and its \lhcb extension, \lhcbdirac~\cite{LHCbDIRAC-SW}. 

\lhcb 's interest in using HPC sites 
is mainly for running \mc (MC) simulation jobs.
MC simulation jobs are, in fact, by far, the largest consumers 
of the \lhcb share of WLCG compute resources
(will be more than 90\% in Run3).
When it comes to distributed computing, 
the \lhcb strategy is 
to use any new compute resources
to run more MC simulation. 
The fraction of non-simulations jobs and CPU, in \lhcb, 
is small enough that we can rely, for them, 
on the currently existing pledged resources.

This paper describes the work that was performed in \lhcb
to be able to run MC simulation jobs on 
the Marconi-A2 HPC facility at CINECA in Bologna, Italy.
An allocation on this supercomputer 
became available to \lhcb in mid-2019 within the context 
of a joint application of the Italian LHC community
for a PRACE grant on this resource,
as described more in detail in another presentation
at this conference~\cite{Boccali2019}.
This work leveraged on the close collaboration
between CINECA and CNAF, 
the Italian Tier-1 site for WLCG,
which is managed by the 
Istituto Nazionale di Fisica Nucleare (INFN)
and is also located in Bologna.

Two different challenges had to be addressed
to integrate the LHCb simulation workflow 
on the Marconi-A2 system,
powered by many-core Intel Knights Landing (KNL) processors,
with limited RAM memory per hardware thread:
first, the \lhcb MC software application,
Gauss~\cite{Clemencic_2011},
had to be re-engineered to use multi-processing (MP)
or multi-threading (MT) to have a lower memory footprint per thread;
second, the \lhcbdirac framework had to 
modified to be able to submit MP or MT jobs on batch queues,
as this was the first time 
these types of jobs were used in \lhcb distributed computing
(even if this is expected to be the norm in the future,
also for other types of workflows such as event reconstruction).

This paper is organized as follows. 
Section~\ref{sec:hpcs} describes the generic challenges
for integrating \lhcb computing workflows on HPC resources.
Section~\ref{sec:marconi} introduces the Marconi-A2 HPC at CINECA. 
Section~\ref{sec:gaussmp} describes 
the work that was done to commission 
multi-processing applications in the Gauss simulation software.
In section \ref{sec:dirac} the \dirac project 
is briefly introduced, with a focus on the capabilities 
of the \dirac Workload Management System. 
Section \ref{sec:distributed} 
gives more details about
the distributed computing challenges and solutions
for \lhcbdirac on Marconi-A2.
Finally, conclusions are given in Section \ref{sec:conclusions}.

\section{Challenges of High Performance Computer systems}
\label{sec:hpcs}

Using HPC facilities,
in \lhcb but more generally for any HEP experiment,
poses two rather distinct types of challenges:
\begin{itemize}
\item Software architecture challenges: 
the compute power of HPC supercomputers 
may come from a range of different processor architectures,
including multi-core and many-core x86 CPUs,
non-x86 CPUs (ARM, Power9) 
and accelerators (GPUs, FPGAs).
For an LHC experiment,
being able to efficiently exploit these resources
may require significant changes to its software applications,
which are generally designed
for the setup of a traditional WLCG worker node,
based on an x86 CPU 
with at least 2 GB RAM available per hardware thread.
In addition, HPCs provide extremely fast inter-node connectivity,
often used for parallel processing using MPI,
while most HEP software applications generally use 
the individual nodes independently of one another,
as if HPCs were just very large clusters.
\item Distributed computing challenges: 
HPC sites usually have 
strict site policies for security reasons
and may therefore be characterized by 
limited or absent external connectivity, 
ad-hoc operating systems,
limited storage on local disks,
restrictive policies for user authentication/authorization
and for batch queues.
This differs from the configuration
of traditional sites in WLCG,
which provide full access to remote services 
like the CernVM File System (CernVM-FS, often abbreviated CVMFS\cite{bib:CVMFS}) for software installation,
uniform operating systems 
and the capability to use user-level virtualization.
\end{itemize}
More generally, unlike WLCG sites 
which provide a relatively uniform computing environment,
HPC centers may differ significantly from one another.
Some HPCs are easier to exploit than others, 
e.g. \lhcb already uses Piz Daint 
at CSCS~\cite{Sciacca_2017}, which looks like 
a traditional Grid site providing a cluster 
of nodes powered by x86 CPUs.
The collaboration of the experiments 
with the local system administrators
and performance experts is in any case essential
to address the specific issues of each HPC center,
and has proved to be mutually beneficial. 

All in all, HPCs are not the most natural fit 
for HEP computing today.
Because of the large amounts of resources
dedicated to scientific computing 
that are currently deployed at HPC centers now,
and of their predicted further increase in the future,
it is however essential that LHC experiments
continue to work on adapting 
their software and computing infrastructures
to be able to efficiently exploit these resources 
in the near future.

\section{Marconi-A2: a KNL based partition of the CINECA supercomputer}
\label{sec:marconi}

Marconi~\cite{marconi} is a supercomputer
at CINECA, 
available for the Italian and European research community.
Currently ranked number 19 in the top500.org list~\cite{top500},
Marconi provides its compute capacity 
through several independent partitions.
The Marconi-A2 partition, 
which has been used for the work described in this paper,
consists of nodes equipped 
with one Xeon Phi 7250 (KNL) at 1.4 GHz, 
with 96 GB of on board RAM. 
This is an x86 many-core CPU with
68 physical cores, supporting 4-way hyperthreading. 
Keeping into account that approximately 10 GB of memory 
are reserved for the O/S,
this means that just over 300 MB of RAM
are available per hardware thread
if all 272 threads are used.
This is much lower 
than the 2 GB (or more) per thread
available at WLCG sites,
motivating the effort to
implement MP and MT approaches in the Gauss software,
as described in the next Section~\ref{sec:gaussmp}.

In January 2020,
most of the A2 partition 
(which included 3600 nodes in 2019) was switched off,
to upgrade Marconi's compute capacity
by replacing the KNLs in Marconi-A2
by the GPUs in a new Marconi100 partition.
Some of the KNLs in A2 are however still available
at the time of writing in March 2020,
and will be used for our \lhcb work
until the end of the granted allocation.
While the efficient exploitation of KNLs
already required some software development effort,
as described in the next section,
it should be noted 
that the work we describe could not have been performed
on the GPU-based partition,
as neither event generation nor detector simulation
are yet possible in the \lhcb software on GPUs.

The default computing environment on Marconi
is also quite different from that normally found at WLCG sites.
Thanks to the excellent collaboration
between the Italian experiment contacts and the 
site managers and sysadmins
at CINECA and CNAF, many essential ad-hoc changes 
were deployed for the LHC experiments~\cite{Boccali2019}.
In particular:
CVMFS mounts and Squids were provided;
external outgoing networking was partially opened at CINECA, 
with routing active to the IP ranges of CERN, FermiLab and CNAF;
the Singularity~\cite{singularity} container 
management tool was deployed;
a HTCondor-CE (the HTCondor Computing Element) was allowed on a CINECA edge node, 
for submitting jobs to the internal SLURM batch system 
connected to Marconi-A2.
The only change that was needed
to allow MP/MT \lhcb MC simulation workflows
on Marconi-A2 was therefore
the implementation of MP/MT job submission
in \lhcbdirac,
as described in Section~\ref{sec:distributed}.

\section{Multi-process \lhcb MC simulation on Marconi-A2: GaussMP}
\label{sec:gaussmp}

The \lhcb software 
for running \mc simulations,
Gauss~\cite{Clemencic_2011},
is used for both event generation~\cite{Belyaev_2011}
and detector simulation, 
the latter using internally 
the Geant4~\cite{AGOSTINELLI2003250} simulation toolkit.
Currently, MC simulation jobs in \lhcb execute
both steps sequentially on the~same worker node, therefore
they need basically no input data (only a configuration file). 
This implies that,
unlike 
workflows such as 
event reconstruction or stripping,
the management of input data files
is not an issue for MC simulation.
In addition, both event generation and detector simulation software applications are compute-intensive, rather than I/O-intensive.
Optimising their performance is mainly
a problem of an efficient use of CPU and memory.

Until recently, \lhcb has only used a
single-process (SP), single-threaded (ST),
version of the Gauss software 
for all of its MC simulation productions.
Only x86 architectures are currently supported, 
although ports to ARM have been worked on.
The typical memory footprint of these applications,
around 1.4 GB~\cite{corti2019},
has so far not been an issue on traditional WLCG nodes,
where 2 GB per thread are available.
On many-core CPUs like the KNL used in Marconi-A2,
however, these workflows are very inefficient,
because the limited memory available per thread
(300 MB if 4 threads are used 
on each of the 68 physical cores of a KNL,
or 600 MB if only 2 threads are used)
effectively limits the maximum number
of SP/ST Gauss instances that can be executed simultaneously,
i.e. the number of KNL threads that can be filled.
To reduce the memory footprint per thread
and be able to fill a larger number of KNL threads,
multi-processing (MP) or multi-threading (MT)
approaches are needed.

In the long term, the \lhcb solution
will be to base its MC simulations
on Gaussino~\cite{muller2019},
a MT implementation of Gauss.
In spite of its very fast recent progress,
however, Gaussino was still not ready 
for meaningful productions or tests
when the Marconi-A2 allocation started.
As a temporary solution,
the software work targeting the Marconi A2 timescales 
focused on the test and commissioning 
of GaussMP, a MP-based version of Gauss.
This leveraged on GaudiMP~\cite{gaudimp2014,nrphd2014},
a MP version of the \lhcb event processing framework Gaudi, 
which~already existed but had never been used in production.

The software work on GaussMP had two main aspects:
extensive functional testing and bug-fixing,
and performance testing and optimization.
Functional testing and bug-fixing 
essentially targeted, and achieved,
a validation of results requiring identical results 
in the SP and MP versions of Gauss,
when simulating the same set of events,
starting from the same random number seeds.
In other words, rather than requiring
a physics validation of results
by comparing physics distributions
within MC statistical errors,
the MP software was validated by requiring
that some event-by-event properties
(numbers of particles and vertices etc.)
should be the same in MP and SP applications.
This implied careful checks 
in both the event generation
and detector simulation steps of the application.

Performance testing and optimization
essentially consisted in running
several identical copies
of an application on a given worker node,
in SP mode or using different MP configurations,
to understand which configuration maximises
the total throughput 
of the entire node,
i.e. the number of events processed 
per unit wall-clock time.
Memory usage was also monitored
to provide an interpretation of throughput results.
The tests were performed both 
on a reference node at CERN,
using a traditional hardware setup
based on two multi-core Haswell CPUs 
with 2 GB per hardware thread
(2x8 physical cores with 2-way hyperthreading and 64 GB RAM),
and on a KNL node from Marconi-A2
(68 physical cores with 4-way hyper-threading and 96 GB RAM).

\begin{figure}[bt]
\begin{center}
\mbox{
\includegraphics[width=12.5cm]{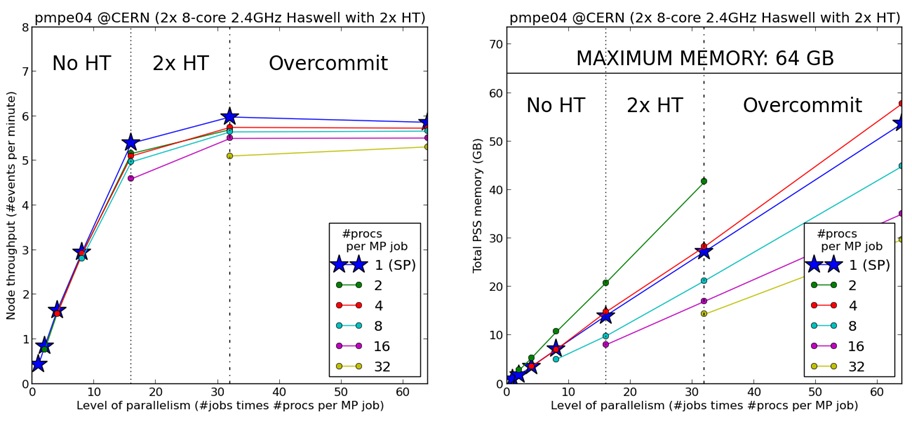} 
}
\end{center}
\vspace*{-7mm}
\caption{Gauss and GaussMP: throughput and memory 
on the reference Haswell node at CERN 
(2x8~physical cores with 2-way hyperthreading and 64 GB RAM).
\vspace*{-5mm}}
\label{fig:haswell}
\end{figure}

\begin{figure}[tb]
\begin{center}
\mbox{
\includegraphics[width=12.5cm]{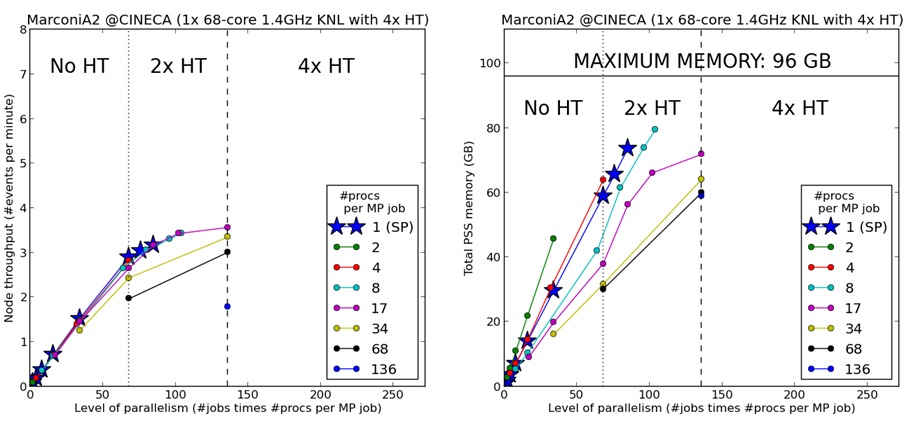} 
}
\end{center}
\vspace*{-7mm}
\caption{Gauss and GaussMP: throughput and memory 
on a Marconi-A2 KNL node at CINECA 
(68~physical cores with 4-way hyper-threading and 96 GB RAM).
\vspace*{-7mm}}
\label{fig:knl}
\end{figure}
The results of these tests
are shown in Fig.~\ref{fig:haswell} 
for the reference node at CERN
and in~Fig.~\ref{fig:knl} 
for the KNL on Marconi-A2.
The memory footprint of the physics process
used for this specific test
(event generation and detector simulation
of $B^+\rightarrow J/\psi\,K^+$ production, including
spillover from minimum bias production in adjacent collisions)
is around 900 MB per process/thread in SP mode.
On the reference node, with 64 GB RAM,
32 instances of an SP application
can be used to fill all 32 threads,
and this is the configuration 
providing \mbox{maximum} throughput
(6.0 events per minute): 
several MP configurations provide similar,
but slightly lower, integrated throughput
(for instance, 5.8 events per minute for 8 instances 
of GaussMP with 4 processes each),
because of the overhead involved 
in the extra processes used by GaussMP.
On the~KNL node, however,
at most 85 SP instances can be launched,
because some processes 
are killed by the out-of-memory monitor
if more instances are launched.
Using GaussMP results in a lower 
memory footprint per hardware thread,
allowing a larger number 
of KNL threads to be filled:
in particular, the maximum throughput on the KNL
is achieved when 8~GaussMP 
application instances 
are executed in parallel, 
each using 17 processes. 
This corresponds to using 2~hardware thread per KNL core 
(i.e. 136 in total), not 4: 
many failures are observed when trying 
to use 4 threads per KNL core (i.e. 272 in total).

The highest GaussMP throughput 
achieved on the KNL
(3.6 events per minute)
is only moderately higher ($\sim$15\%) 
than that achieved using SP Gauss
(3.2 events per minute), 
because the forking strategy used
does not optimize the use of copy-on-write
to minimize the memory footprint.
Currently, new worker processes are forked 
after job initialization but
before the first event~\cite{gaudimp2014,nrphd2014},
where the magnetic field map is read from disk~\cite{bmklhcb2019};
forking workers after 
processing the first event
would make it possible 
to share a larger amount of memory across workers
and reduce the overall memory footprint,
as recently demonstrated by ATLAS
in their new AthenaMP forking strategy~\cite{elms}.
Looking forward, however, 
\lhcb software efforts in the simulation area 
will focus on Gaussino, the long-term MT solution,
rather than on GaussMP, the temporary MP solution.

It is also interesting to note that,
in absolute terms, 
the throughput per thread
achieved in the configuration
maximising the total node throughput
is a factor 7 lower on the KNL
(0.026 events per minute per core,
for 3.6 events per minute on 136 threads)
than on the reference Haswell node
(0.188 events per minute per thread,
for 6.0 events per minute on 32 threads).
This can only be partly explained
in terms of the lower clock speed
of the KNL cores,
and is probably also due to the memory access patterns of the Gauss application
on the two architectures,
but no specific studies have been performed to understand this better.

\section{The \dirac project}
\label{sec:dirac}

\dirac~\cite{DIRAC-SW} is a software framework 
that enables communities to interact 
with distributed computing resources. 
It builds a layer between users and resources, 
hiding diversities across computing, storage, 
catalog, and queuing resources. 
\dirac has been adopted by several HEP 
and non-HEP experiment communities \cite{DIRAC-HEP-CHEP2016}, 
with different goals, intents, resources and workflows: 
it is experiment agnostic, extensible, and flexible \cite{GPUDIRAC}. 
\lhcb uses \dirac for managing 
all its distributed computing activities.
\dirac is an open source project, 
which was started around 2002 as an \lhcb project. 
Following interest of adoption from other communities 
its code was made available under open licence in 2009. 
Now, it is hosted on
GitHub~\cite{dirac-github}
and is released under the GPLv3 license. 

The \dirac Workload Management System (WMS) 
is in charge of exploiting distributed computing resources. 
In other words, it manages jobs, 
and pilot jobs~\cite{DIRAC-PILOTS_2016} 
(from here on simply called "pilots").
The emergence of new  distributed computing resources 
(private and commercial clouds, 
High Performance Computing clusters, volunteer computing, etc) 
changed the traditional landscape of computing 
for offline processing. 
It is therefore crucial to provide 
a very versatile and flexible system 
for handling distributed computing 
(production and user data analysis). 
If we restrict for a moment our vision to LHC experiments, 
and we analyze the amount of CPU cycles they used in the last year,
we can notice that all of them have consumed 
more CPU-hours than those official reserved (pledged) 
to them by WLCG.  
Each community found ways to exploit 
``opportunistic'', i.e. non-pledged, 
compute resources (CPUs, or even GPUs).
Such resources may be private to the experiment 
(e.g. the ``online'' computing farm - 
often simply called ``High Level Trigger'' farm) or public; 
resources may sometimes be donated free of charge, 
like in the case of volunteer computing,~or not, 
like public commercial cloud providers. 
Integrating non-grid resources is common to~all communities 
that have been using WLCG in the past, and still do. 
Communities that use~\dirac~want to exploit 
all possible CPU or GPU cycles. 
Software products like \dirac aim to make this easy, 
and the \dirac pilot is the \emph{federator} 
of each and every computing resource. 
With \dirac, transparent access to the computing resources
is realized by implementing the pilot~model. 

\section{Distributed computing challenges on Marconi-A2: fat nodes}
\label{sec:distributed}

The integration of new compute resources 
into the \lhcb distributed computing framework, based on DIRAC,
is generally an easy task when: 
first, worker nodes (WNs) have outbound network connectivity; 
second, the \lhcb CVMFS endpoints are mounted on the WNs; 
and, third, the WN O/S is an SLC6 (Scientific Linux CERN 6) 
or CC7 (CERN CentOS 7) compatible flavor, 
or Singularity containers are available. 
As discussed in Sec.~\ref{sec:hpcs},
none of these conditions would normally be satisfied on Marconi,
but all three were eventually met on Marconi-A2
specifically for the LHC experiments,
thanks to the good collaboration 
between the experiment contacts and the site managers.

\begin{figure}
    \centering
    \includegraphics[width=5cm]{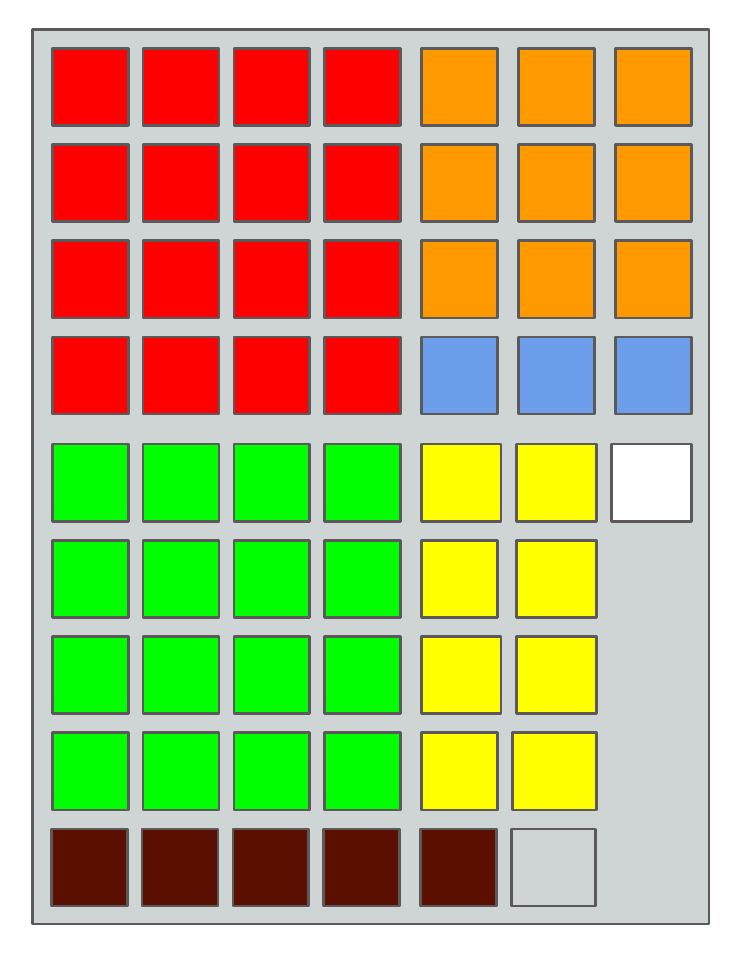}
    \caption{Allocating jobs within a fat node: the DIRAC pilot, when using the Pool inner Computing Element, realizes a de-facto partitioning of the node, and parallel jobs matching. The figure above shows a theoretical allocation of jobs to logical processors, with each box representing a logical processor. Each color represents a different application.
    \vspace*{-6mm}}
    \label{fig:fat}
\end{figure}

\begin{figure}
    \centering
    \def\svgwidth{\columnwidth}
    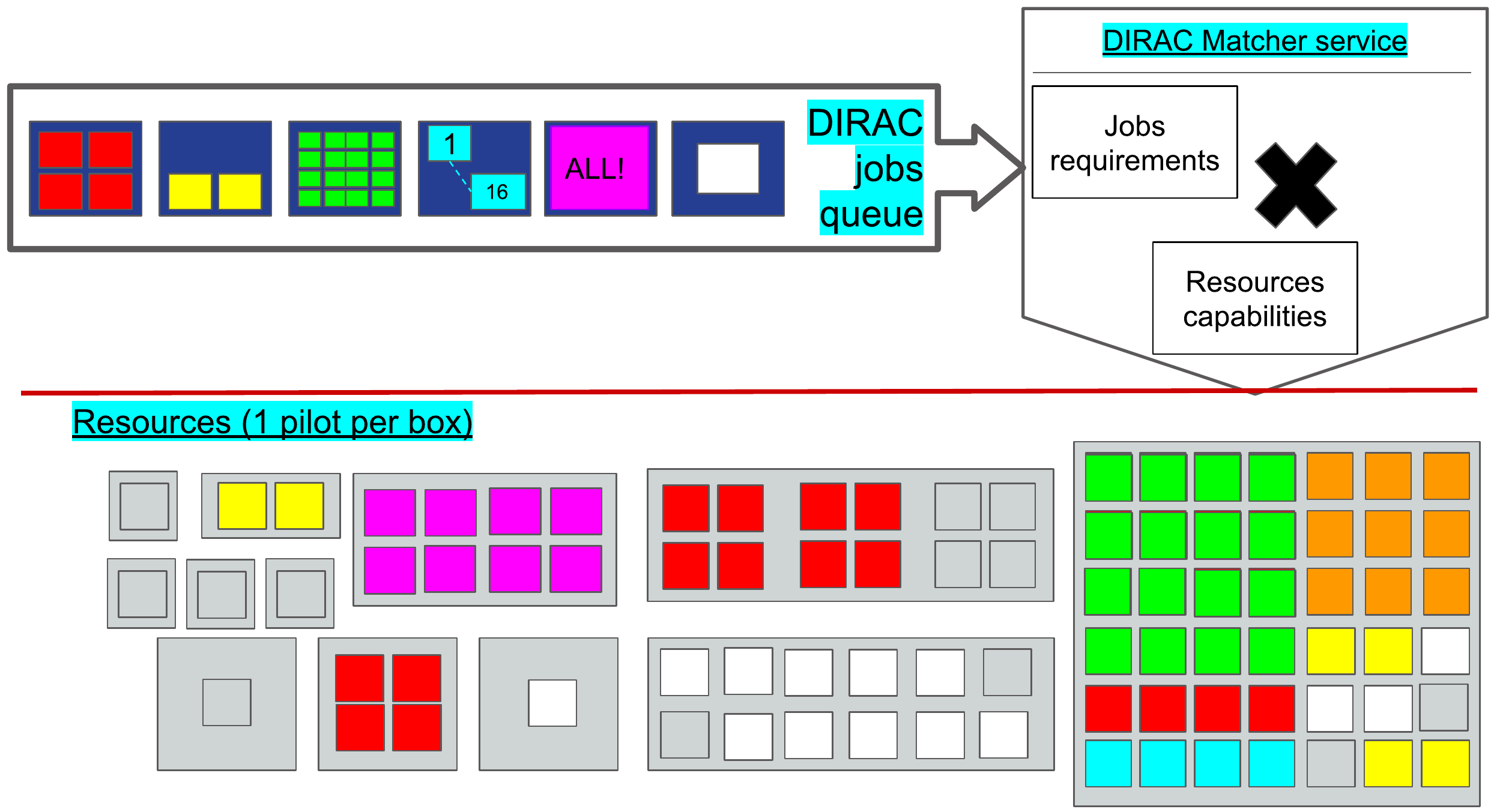
    \caption{Job matching: matching job requirements to computing resources.
    \vspace*{-6mm}}
    \label{fig:match}
\end{figure}

As a consequence,
on the distributed computing side, 
the main challenge \lhcb had to address 
was that each job slot provided by the HTCondorCE 
represents a whole KNL node, 
with 68 physical cores and up to 4 hardware threads per core,
i.e. a total of {\tt nproc}=272 logical processors
(assuming that 4-way hyper-threading is enabled).
Rather than implementing a quick ad-hoc solution for Marconi-A2, 
this was addressed in \dirac by developing 
a generic mechanism for managing "fat nodes",
as shown schematically 
in Figs.~\ref{fig:fat} and~\ref{fig:match}.
In~\dirac terminology, 
this problem, of how to 
subdivide a fat node and allocate its resources to several jobs,
is simply called ``matching''.
It is worth noting that this was never done in \lhcb before, 
as only jobs running SP/ST software workloads and
using a single logical processor were used. 

For a proper job allocation, 
\dirac needs to “partition” the node 
for optimal memory and throughput 
(and maybe only use a subset of the logical processors). 
For this to happen, 
we have developed the \emph{Pool} ``inner'' Computing Element, 
with which it is possible to execute parallel jobs matching. 
In the following, 
by ``processor'' we mean a ``logical processor''
(whose number is {\tt nproc} in total);
by ``single-processor'' jobs we mean 
single-threaded, single-process software application workloads,
requiring a single logical processor,
while by ``multi-processor'' job we mean
a software application workload that uses more than one logical processor,
whether the application is implemented using a multi-process approach
(like GaussMP)
or a multi-threaded approach
(like Gaussino), or a combination of both.

From a user's perspective, 
it is possible to describe the jobs precisely enough 
to satisfy all use cases below:
\begin{itemize}
    \item certain jobs may be able to run 
    only in single-processor mode
    \item certain jobs may be able to run 
    only in multi-processor mode 
    (i.e., they need at least 2 logical processors)
    \item certain multi-processor jobs 
    may need a fixed amount of logical processors
    (in our specific case on Marconi-A2,
    we chose to submit only GaussMP jobs,
    at most 8 simultaneously,
    using 17 logical processors per job,
    to maximize the whole node throughput)
    \item certain jobs may be able to run both 
    in single-processor or multi-processor mode, 
    depending on what is possible on the WN/Queue/CE
    \item for certain jobs we may want to specify 
    a maximum number of processors to use
\end{itemize}

At the same time, 
from a resource provider's perspective, 
it is possible to describe CEs and Queues precisely enough 
to satisfy all use cases below:
\begin{itemize}
    \item may give their users the possibility 
    to run on their resources:
    \begin{itemize}
        \item only single processor jobs
        \item both single and multi processor jobs
    \end{itemize}{}
    \item may ask their users to distinguish clearly 
    between single and multi processor jobs
    \item may need to know the exact number of processors 
    a job is requesting
    \item may ask for only ``wholeNode'' jobs
\end{itemize}

\section{Summary and outlook}
\label{sec:conclusions}

In summary,
both of the challenges involved 
in the integration of LHCb MC simulation workflows 
on the Marconi-A2 HPC 
have been addressed:
a multi-process version of the Gauss software framework
with reduced memory footprint per thread 
has been commissioned,
and the functionality of managing fat nodes
has been added to the LHCdDIRAC 
distributed computing framework
and has been successfully tested
in a dedicated certification environment.
At the time of writing in March 2020,
however, the new \lhcbdirac functionality
has not yet been deployed in production, 
because of the timescales involved 
in the LHCb software release process. 
This is the reason why no results 
of production use of Marconi A2 by LHCb 
for MC simulation using GaussMP are shown in this paper.
As soon as the new \lhcbdirac is released
within the next few weeks, however,
the remaining LHCb allocation on the Marconi-A2 KNLs
will be used to launch the first production jobs of MC simulation,
using software workflows based on GaussMP,
as well as using Gaussino if available on time.

More generally,
this effort at integrating a new HPC resource
into the LHCb software and computing 
was extremely valuable.
On the software application side,
it was useful to highlight
some of the challenges ahead
in the use of non-traditional compute architectures
(which may well be GPUs in the not-so distant future).
On the distributed computing side,
it was useful to pave the way
to the more routine use 
of multi-threaded software applications on the grid,
which will soon become the norm.
Last but not least, the collaboration
with the other LHC experiments
and with the local site managers and sysadmins at CINECA and CNAF
was an essential ingredient of this effort,
and a pleasant and fruitful experience 
for which we thank them, and that 
we look forward to repeating in the future.


\begin{thebibliography}{} 

\bibitem{DIRAC-SW} 
  F. Stagni et al.,
  {\em DIRACGrid/DIRAC} (2018).
  \doi{10.5281/zenodo.1451647}

\bibitem{LHCbDIRAC-SW}
  LHCb Coll., 
  {\em LHCbDIRAC} (2018). 
  \doi{10.5281/zenodo.1451768}

\bibitem{Boccali2019} 
  T. Boccali et al., 
  {\em Extension of the INFN Tier-1 on a HPC system}, 
  to appear in Proc. CHEP2019, Adelaide (2019). \url{https://indico.cern.ch/event/773049/contributions/3474805}
  
\bibitem{Clemencic_2011}
  M. Clemencic et al.,
  {\em The LHCb Simulation Application, Gauss: Design, Evolution and Experience},
  Proc. CHEP2010, Taipei,
  J. Phys. Conf. Ser. \textbf{331}, 032023 (2011).
  \doi{10.1088/1742-6596/331/3/032023}
  
\bibitem{bib:CVMFS}
  J. Blomer et al.,
  {\em Distributing LHC application software and conditions databases using the CernVM file system},
  Proc. CHEP2010, Taipei,
  J. Phys. Conf. Ser. \textbf{331}, 042003 (2011).
  \doi{10.1088/1742-6596/331/4/042003}

\bibitem{Sciacca_2017}
  F. G. Sciacca, S. Haug et al.,
  {\em ATLAS and LHC computing on CRAY},
  Proc. CHEP2016, San Francisco,
  J. Phys. Conf. Ser. \textbf{898}, 082004 (2017).
  \doi{10.1088/1742-6596/898/8/082004}

\bibitem{marconi}
  Marconi at CINECA, 
  \url{http://www.hpc.cineca.it/hardware/marconi}
  
\bibitem{top500}
  top500 rankings as of November 2019, 
  \url{https://www.top500.org/lists/2019/11}

\bibitem{singularity}
  G. M. Kurtzer, V. Sochat, M. W. Bauer, 
  {\em Singularity: Scientific containers for mobility of compute}, 
  PLoS ONE \textbf{12}, e0177459 (2017). 
  \doi{10.1371/journal.pone.0177459}

\bibitem{Belyaev_2011}
  I. Belyaev et al.,
  {\em Handling of the generation of primary events in Gauss, the LHCb simulation framework},
  Proc. CHEP2010, Taipei,
  J. Phys. Conf. Ser. \textbf{331}, 032047 (2011).
  \doi{10.1088/1742-6596/331/3/032047}

\bibitem{AGOSTINELLI2003250} 
  S. Agostinelli et al., 
  {\em Geant4 --- a simulation toolkit},
  NIM A \textbf{506}, 250 (2003).
  \doi{10.1016/S0168-9002(03)01368-8}

\bibitem{corti2019}
  G. Corti et al., 
  {\em Computing performance of the LHCb simulation},
  24th Geant4 Collaboration Meeting, JLAB (2019). 
  \url{https://indico.cern.ch/event/825306/contributions/3565311}

\bibitem{muller2019} 
  D. Muller, 
  {\em Gaussino - a Gaudi-based core simulation framework}, 
  to appear in Proc. CHEP2019, Adelaide (2019).
  \url{https://indico.cern.ch/event/773049/contributions/3474740}

\bibitem{gaudimp2014}
  N. Rauschmayr, A. Streit, 
  {\em Preparing the Gaudi framework and the DIRAC WMS for multicore job submission},
  Proc. CHEP2013, Amsterdam,
  J. Phys. Conf. Ser. \textbf{513}, 052029 (2014).
  \doi{10.1088/1742-6596/513/5/052029}

\bibitem{nrphd2014}
  N. Rauschmayr, 
  {\em Optimisation of LHCb applications for multi- and manycore job submission}, 
  CERN-THESIS-2014-242 (2014).
  \url{https://cds.cern.ch/record/1985236}

\bibitem{bmklhcb2019}
  A. Valassi, S. Muralidharan, 
  {\em Trident analysis of the LHCb GEN/SIM workload from the
benchmarking suite}, 
  System performance modelling WG meeting, CERN (2019).
  \url{https://indico.cern.ch/event/772026}
  
\bibitem{elms}
  J.\,Elmsheuser et al.,
  \mbox{\em ATLAS\,Grid\,Workflow\,Performance\,Optimization},\,Proc.\,CHEP2018,
  Sofia,\,EPJ\,Web\,of\,Conf.\,\textbf{214},\,03021\,(2019).
  \doi{10.1051/epjconf/201921403021}

\bibitem{DIRAC-HEP-CHEP2016}
  F. Stagni et al., 
  {\em DIRAC in Large Particle Physics Experiments},
  Proc. CHEP2016, San Francisco,
  J. Phys. Conf. Ser. \textbf{898}, 092020 (2017).
  \doi{10.1088/1742-6596/898/9/092020}

\bibitem{GPUDIRAC}
  S. Camarasu-Pop et al., 
  {\em Exploiting GPUs on distributed infrastructures for medical
imaging applications with VIP and DIRAC},
  Proc. MIPRO2019, Opatija (2019).
  \doi{10.23919/mipro.2019.8757075}

\bibitem{dirac-github}
  DIRAC project GitHub repository, 
  \url{https://github.com/DIRACGrid}

\bibitem{DIRAC-PILOTS_2016} 
  F. Stagni et al.,
  {\em DIRAC universal pilots},
  Proc. CHEP2016, San Francisco,
  J. Phys. Conf. Ser. \textbf{898}, 092024 (2017).
  \doi{10.1088/1742-6596/898/9/092024}

\end{thebibliography}
\end{document}